\documentclass[twocolumn,prl,aps,superbib,tightenlines,floatfix,superscriptaddress]{revtex4}
\usepackage[T1]{fontenc}
\usepackage[latin9]{inputenc}
\usepackage{amsmath}
\usepackage{graphicx}
\usepackage{amssymb}
\begin{document}
\title{Chaotic memristor}
\author{T. Driscoll}
  \altaffiliation {Also at: Center for Metamaterials and Integrated Plasmonics, Duke University.}
  \affiliation {Department of Physics, University of California, San Diego, 9500 Gilman Drive, La Jolla, California 92093, USA}
\author{Y. V. Pershin}
  \affiliation {Department of Physics and Astronomy and USC Nanocenter, University of South Carolina, Columbia, South Carolina 29208, USA}
\author{D. N. Basov}
  \affiliation {Department of Physics, University of California, San Diego, 9500 Gilman Drive, La Jolla, California 92093, USA}
\author{M. Di Ventra}
  \affiliation {Department of Physics, University of California, San Diego, 9500 Gilman Drive, La Jolla, California 92093, USA}
\date{Accepted to Applied Physics A, 16 November 2010}
%
%
%

\begin{abstract}
We suggest and experimentally demonstrate a chaotic memory resistor (memristor). The core of our approach is to use a
resistive system whose equations of motion for its internal state variables are similar to those describing a particle in a
multi-well potential. Using a memristor emulator, the chaotic memristor is realized and its chaotic properties are measured.
A Poincar\'{e} plot showing chaos is presented for a simple nonautonomous circuit involving only a voltage source directly
connected in series to a memristor and a standard resistor.  We also explore theoretically some details of this system, plotting the attractor and calculating Lyapunov exponents.  The multi-well potential used resembles that of many nanoscale memristive devices, suggesting the possibility of chaotic dynamics in other existing memristive systems.
\end{abstract}

\maketitle

\section{Introduction}
\label{intro}


The class of memory circuit elements consisting of memristors, memcapacitors and meminductors \cite{diventra09a,pershin10f} is a relatively
new paradigm based on the understanding that many physical realizations of basic circuit elements (resistors, capacitors and
indicators) may involve an intrinsic memory mechanism, at least on certain time scales. Currently, a lot of attention
is devoted to such elements and their potential applications - with a primary focus on memristive elements \cite{chua71a,chua76a,strukov08a}.
Physical manifestations of memristors range from thermistors \cite{sapoff63a} to complex oxide
\cite{yang08a,Jo09a,Gergel09a,driscoll09b,driscoll09a} and spintronic \cite{pershin08a,pershin09a,wang10a} materials. A number of
possible intriguing applications of memristors have already been discussed including digital memory \cite{Kuekes05a,Lehtonen09a,Strukov09c}, neuromorphic systems \cite{snider08a,pershin09b,pershin10c}, adaptive filters \cite{driscoll10a,Muthuswamy09a}, tunable and reconfigurable metamaterials \cite{driscoll09a}, and others \cite{pershin09d,jo10a,Itoh10a}.

By definition \cite{chua76a,diventra09a},  an $n$th-order voltage-controlled memristive system is described by the relations
\begin{eqnarray}
I(t)&=&R^{-1}_M\left(x,V_M,t \right)V_M(t) \label{Condeq1}\\
\dot{x}&=&f\left( x,V_M,t\right) \label{Condeq2}
\end{eqnarray}
where $V_M(t)$ and $I(t)$ denote the voltage across and current through the device, $R_M$ is a scalar called the {\em memristance},
$x$ is a vector representing $n$ internal state variables, and $f$ is an $n$-component vector function. Following present convention,
we will call a device described by Eqs. (\ref{Condeq1})-(\ref{Condeq2}) a memristor, even though the latter was originally defined when
$R_M$ depends on charge or flux (time integral of voltage) only~\cite{chua71a}.

In this paper, we consider a specific type of second-order memristor described by Eq. (\ref{Condeq2}) of the form
\begin{equation}\label{eqn_motion}
\frac{\textnormal{d}}{\textnormal{d}t } \left[
\begin{array}{c} x \\ \\ \dot{x} \end{array} \right]= \left[
\begin{array}{c} \dot{x} \\ \\
-\frac{1}{m}\frac{\partial U }{\partial x}-\gamma\,\dot{x}+\frac{F\left(V_M\right)}{m}
\end{array} \right],
\end{equation}
where $x$ and $\dot{x}$ are internal state variables, $m$ is an effective mass associated with the internal state of the device,
$U(x)$ is a multi-well potential such that $U(\pm \infty)=\infty$ (see schematic in Fig.~\ref{FigCircuit}), $\gamma$ is a damping coefficient and $F$ is a driving force that
depends on the voltage applied to the memristor.

It is important to note that Eq. (\ref{eqn_motion}) describes the classical dynamics of an
effective particle in a multi-well potential and has a close connection to physical processes in certain nanoscale
memristors \cite{strukov09a}. In particular, oxygen vacancy migration in TiO$_2$ memristors can be described as a hopping process
between potential minima \cite{strukov09a}. Moreover, a strained elastic membrane memcapacitor (memory capacitor)
\cite{Martinez10a} is also described by equations similar to Eq. (\ref{eqn_motion}).  These connections give physical grounding to our choice of
memristive function.

An important consequence of Eq. (\ref{eqn_motion}) is the existence of a chaotic regime within a certain range of
parameters.  This approach to chaos with memristors is unique and radically different from previous proposals
\cite{Itoh08a,Muthuswamy09a,Bao10b,bao10a} that are based on modified Chua's circuits \cite{Matsumoto84a,chuabook} with {\it active}
elements. We emphasize also that our realization of chaos results from the simplest single-element nonautonomous chaotic circuit consisting of the chaotic memristor directly connected to a voltage source.  The presence of chaos in such a simple system certainly has potential applications in cryptography and communications, as well as other areas.  It is also converserly true however, that chaotic behavior can be a detriment in memory or computation systems - areas where memristors are currently pushing towards prototype devices.  Thus, understanding the origins of chaos in memristive systems may be critical to avoid or elminate chaos.

The layout of this paper is as follows:  In section 2, the chaotic regime is demonstrated experimentally.  For the experiment, we have built a memristor emulator \cite{pershin09d,pershin10b,pershin10c}, namely a simple circuit that can emulate essentially any memristive function $f$, and has been used by two of the present authors (YVP and MD) to build memcapacitor and meminductor emulators~\cite{pershin10b}, discuss learning and associative memory~\cite{pershin10c}, build programmable analog circuits~\cite{pershin09d}, and even perform logic and arithmetic operations~\cite{pershinsubmitted}. In this work we have implemented the double-well potential of Eq. (\ref{eqn_motion}).
Applying ac-pulse sequences, we explicitly show chaotic behavior within a certain range of applied voltage amplitudes.  Then, in section 3, we connect our experiment to a theoretical investigation of this system, including exploration of the phase-space and Lyapunov exponents for such double-well potentials. Our main results and conclusions are summarized in section \ref{sec:4}.

\section{Experimental demonstration of chaos}
\label{sec:2}

\begin{figure}[t]
\centering
\includegraphics[width=7cm]{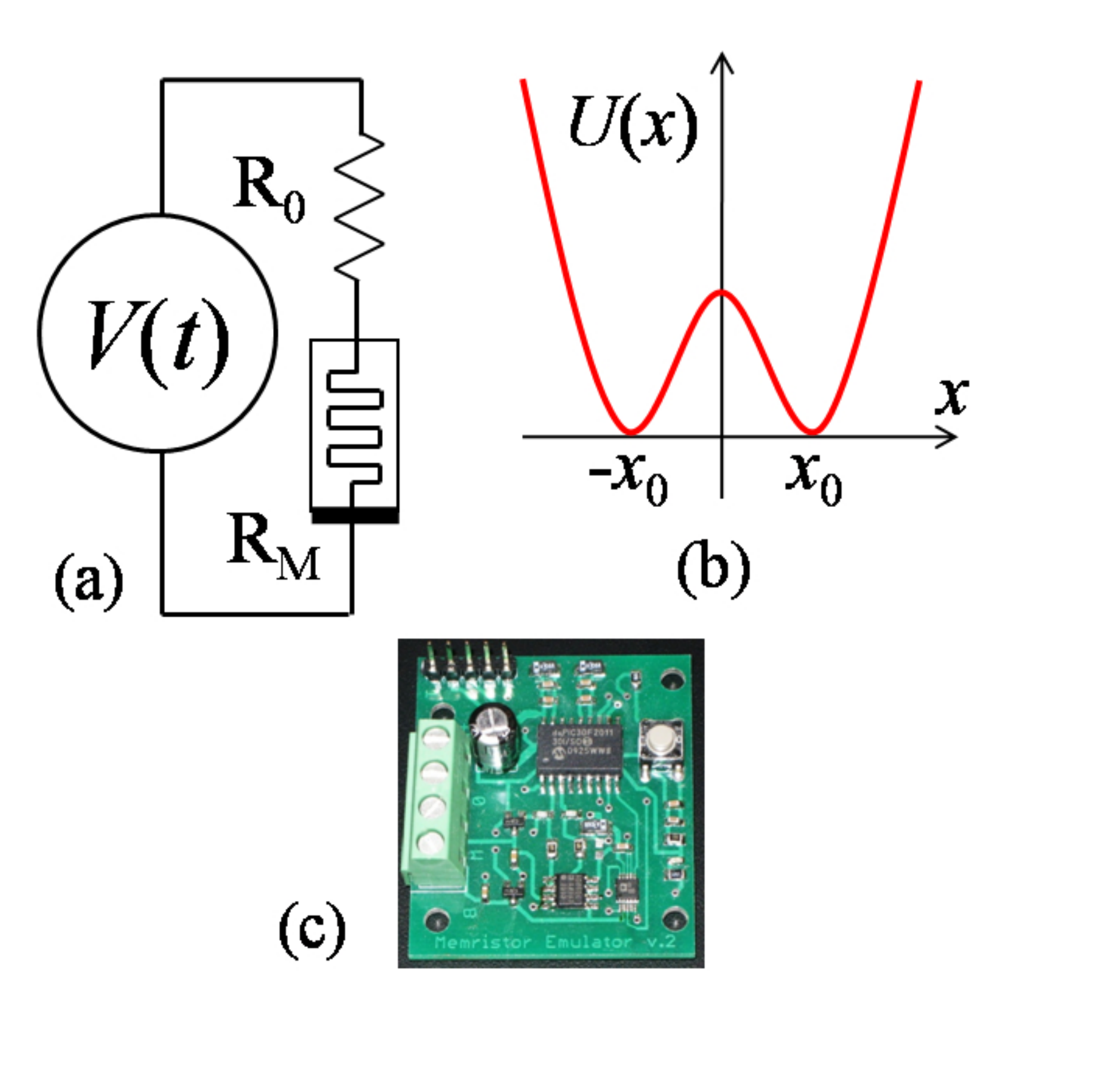}
\caption{(a) Electronic circuit used in chaos measurements consisting
of the programmable voltage source $V(t)$, 510 Ohms resistor R$_0$ and
memristor M. The memristor specifications are given in the text.
(b) Schematic of the double-well potential governing the memristor.
(c) Optical photograph of memristor emulator.}
\label{FigCircuit}       
\end{figure}

For our experiment, the simple electronic circuit shown in Fig. \ref{FigCircuit}(a) was constructed. A programmable voltage source $V(t)$ is connected in series with
the memristor $M$ and a current-sensing resistor $R_0$.  A memristor emulator \cite{pershin09d,pershin10b,pershin10c} whose circuit image is shown in Fig. \ref{FigCircuit}(c) was employed for the memristor.
At the electronic component level, the memristor emulator consists of a digital potentiometer, analog-to-digital converter (ADC) unit
and control unit (microcontroller). The control unit reads the digital code from the ADC (corresponding to the applied voltage) and
generates (and writes) a code for the digital potentiometer according to the predefined algorithm. For the case of the double-well potential we have re-written Eq. (\ref{eqn_motion}) in the form
\begin{equation}\label{programmed_eq}
\ddot{x}=-c_1\dot{x}-c_2x\left(x^2-1 \right)+c_3V_M(t).
\end{equation}
where $c_1=2$ s$^{-1}$, $c_2=100$ s$^{-2}$ and
$c_3=30$ V$^{-1}$s$^{-2}$. The function $R_M(x)$ was then chosen as
$R_M=(4875+1560x)$ Ohms. According to Eq. (\ref{programmed_eq}) at $V_M=0$ the potential minima $\pm x_0$ are located
at $\pm 1$.

We also make use of a control memristor, programming our emulator with a standard threshold-type model \cite{pershin09b}
\begin{eqnarray}
I(t)&=&x^{-1}V_M(t), \label{control_eqZ} \\
\dot{x}&=&-\beta V_M + 0.5 \beta (\vert V_M + V_T \vert - \vert V_M - V_T\vert )\label{control_eq}
\end{eqnarray}
with the parameters $\beta$=62k$\Omega$/(V$\cdot$s), $V_T$=0.625 V, and the memristance can change only between
$R_1=1$k$\Omega$ and $R_2=10$k$\Omega$. In the actual microcontroller's software
implementation, the value of $x$  is monitored at each time step
and in the situations when $x<R_1$ or $x>R_2$, it is set equal to
$R_1$ or $R_2$, respectively.

In this circuit, the applied voltage and the voltage across $R_0$ are both measured at 800 Hz, in hardware-timed synchronization with the update-rate of $V(t)$. The sequence of the applied voltage $V(t)$ is as follows: starting from zero applied voltage, we apply a large negative voltage (-2.25V) for 2 seconds, followed by zero voltage for 1 second.  This acts as a reset, ensuring we know the initial state of the memristor.  After this reset, a sinusoidal voltage of 2 Hz is applied with a peak amplitude $V_{app}$.  This is applied for 10 seconds for the control memristor, and 20 seconds for the chaotic memristor.  After this, the voltage is returned to zero, and the system is allowed to rest for 5 seconds.  Then the complete sequence (including a reset each time) is repeated for increasing values of $V_{app}$ between 0.1 V and 2.9 V with 1000 steps.  It is important to include a reset in each sequence, as the memristor retains information about previous voltages applied to it.  By utilizing such a reset we can insure that initial conditions are known, a very important point for chaotic systems.

\begin{figure}[tbh] 
\centering
\includegraphics[width=8.5cm]{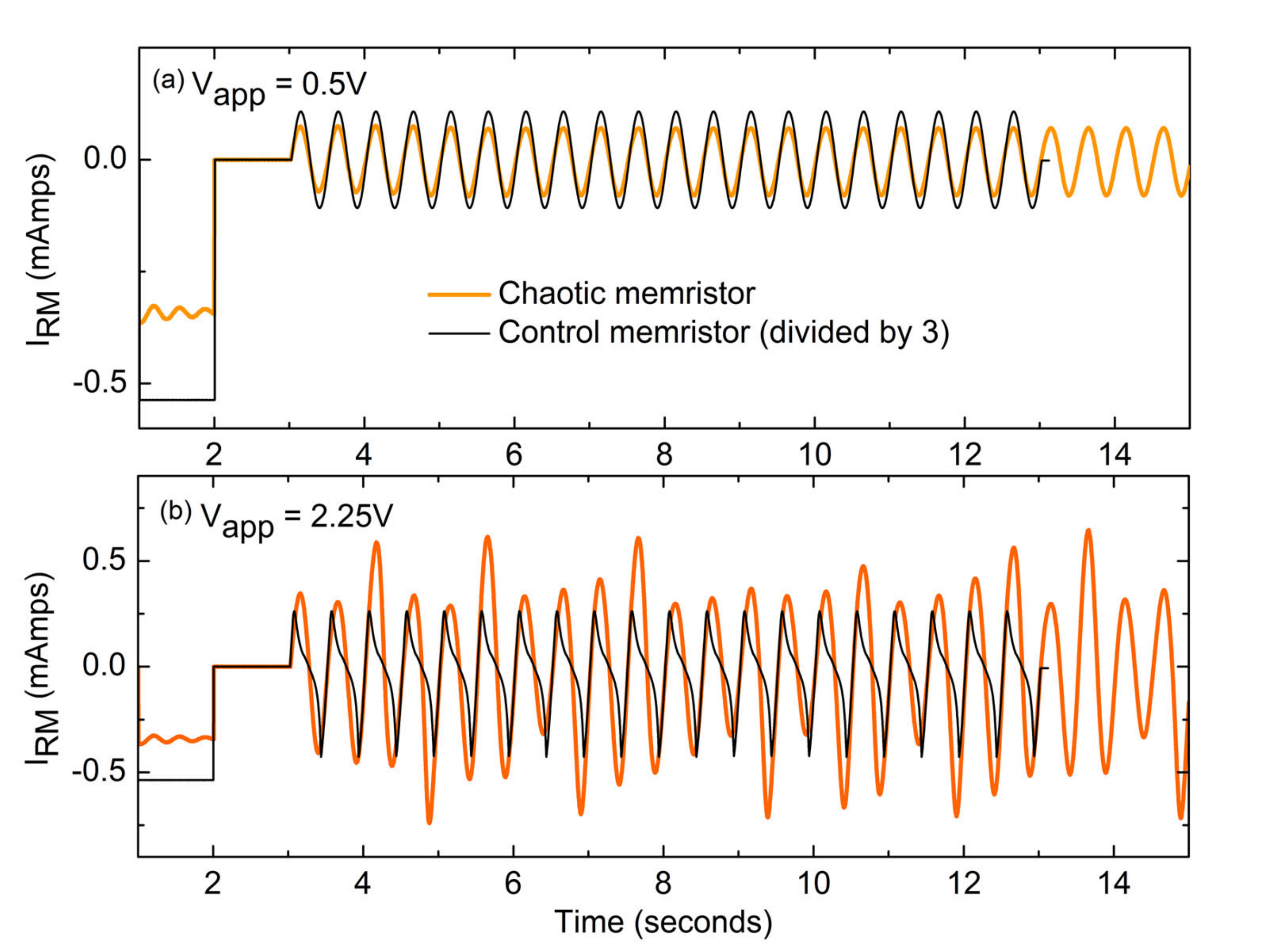}
\caption{Traces showing current through the memristor for two cases: a control memristor and our chaotic memristor.
The chaotic memristor operates according to Eqs.~(\ref{Condeq1},\ref{programmed_eq}) with $R_M$ defined below Eq. (\ref{programmed_eq}).  The control memristor operates according
to Eqs.~(\ref{control_eqZ},\ref{control_eq}).
Traces are shown for two voltages: (a) The applied voltage amplitude is below the threshold where both systems behave linearly. (b) The applied voltage amplitude is above threshold.  In the above-threshold region for our chaotic memristor, notice that the amplitude of any peak follows no pattern, while the control memristor peaks remain constant even though it follows an un-harmonic wave-shape.
In each case the initial low voltage is a ``reset" procedure to verify the state of the memristor is known at the onset of applied oscillations.}
\label{FigIV}       
\end{figure}

From the measured $V_{R_0}$ the instantaneous current $I(t)$ through the circuit can be calculated at any time.  As indicated in Eq. (\ref{Condeq1}), $I(t)$ is a critical parameter for understanding the behavior of $R_M$.  Figure \ref{FigIV} plots such current for four important cases.  In Fig. \ref{FigIV}(a), current traces are plotted for our control memristor (black) and chaotic memristor (orange) for $V_{app}=0.5$ V, which is below the threshold voltage for both systems.  For such sub-threshold voltages, the response of both systems is nearly linear, and we observe periodic sinusoidal currents.  In Fig. \ref{FigIV}(b) we repeat this for $V_{app}=2.25V$, which is above threshold voltage.  In this case we observe the control memristor (black) circuit exhibits very nonlinear currents.  However, despite the un-harmonic shape, the amplitudes of the current peaks (both positive and negative) are nearly constant and predictable from one to the next.  In contrast, for the chaotic memristor the current-peak amplitude from one cycle to the next can be very different, and change in an unpredictable (pattern-less) manner.  This is a fingerprint of chaotic behavior.

\begin{figure}[tbh] 
\centering
\includegraphics[width=8.5cm]{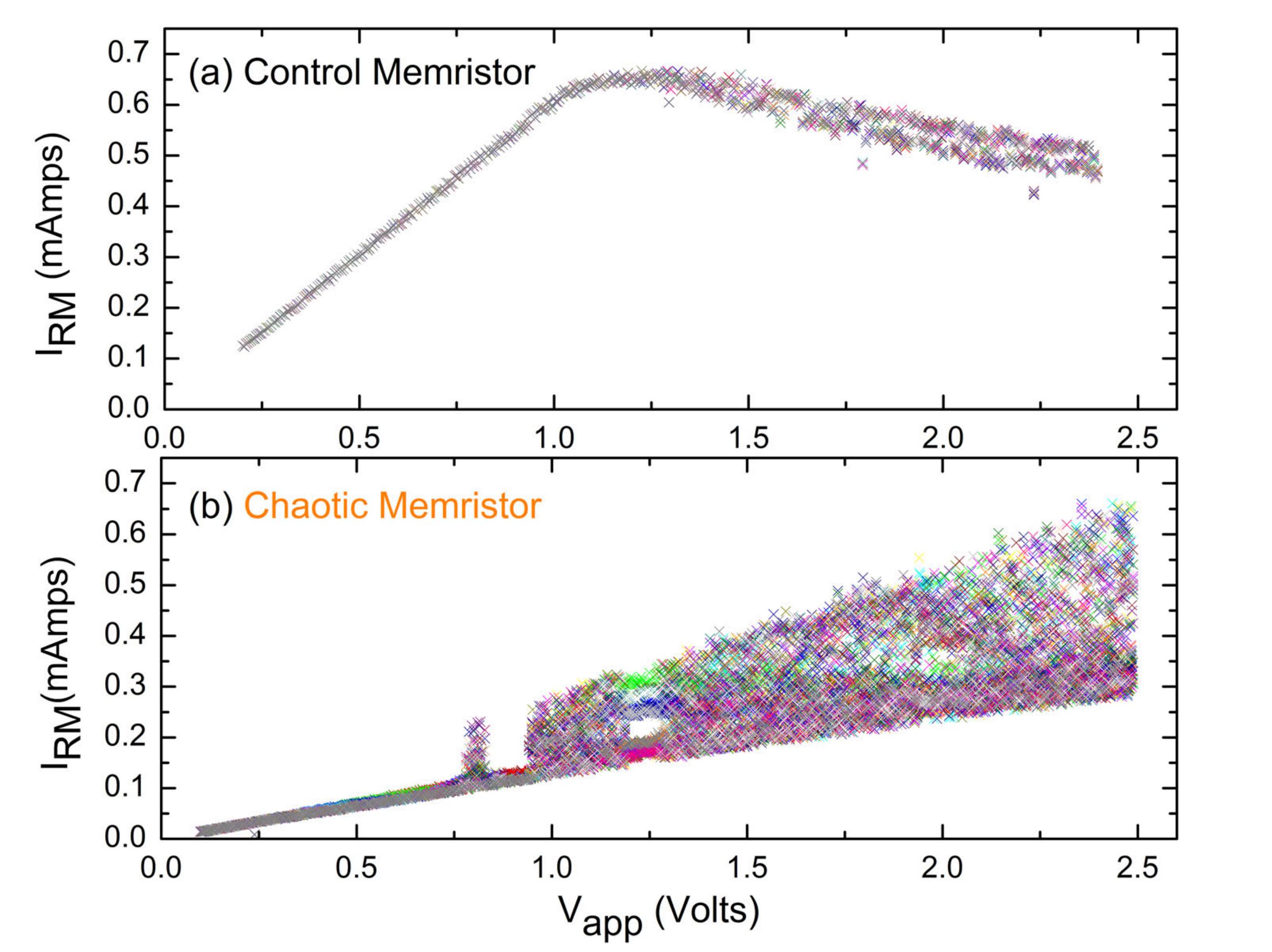}
\caption{Poincar\'{e} graphs constructed from the oscillation peaks from the $I-V$ datasets (such as those in \ref{FigIV}).  For the control memristor (a), the graph shows no chaotic behavior anywhere.  For the chaos memristor (b), above a certain threshold
(about 1 V) the behavior is chaotic.  The threshold at which this chaos onset is determined by the potential barrier between the
double well.  The colors used for sequential points follows a repeating rainbow colormap.  This further helps to show the lack of any pattern, as no color pattern/structure is evident. }
\label{FigPoincare}       
\end{figure}

In order to better illustrate this chaotic behavior, we construct Poincar\'{e} plots from the complete $I(t)$ dataset.  To do this, the amplitude of $I(t)$ is extracted at periodic times of the driving frequency, selecting the positive peak of each oscillation.  All peaks in each $V_{app}$ dataset are plotted as a function of the $V_{app}$, except the first 3 periods from each $V_{app}$ are omitted to exclude any transient effects.  This procedure gives 37 values for the chaotic memristor and 17 for control memristor for each $V_{app}$.  These are plotted in Fig. \ref{FigPoincare}.

Looking at the Poincar\'{e} plots, we see very different behavior between the control memristor and the chaotic memristor.  In the low voltage region ($<$1 V) of the control memristor (Fig. \ref{FigPoincare}a), the 17 points overlap almost exactly, creating a linear response.  Above 1 V, two narrowly split branches emerge.  Most probably, these branches are a result of the finite resistance descritization in our memristor emulator.  For the rapid $\dot{x}$ above threshold, errors stemming from the relatively slow internal update rate of the emulator causes the resistance value to jitter two neighboring digital values.  This small effect (only $\sim$1\% changes in R) comes only from our emulator implimentation and not the memristive dynamics itself.  The negative differential resistance also stems from the function Eq. (\ref{control_eq}) of our control memristor \cite{pershin09b}.  The Poincar\'{e} plot for the chaotic memristor looks very different.  At low voltages, it also behaves as a normal linear resistor.  However, above a threshold voltage of about 1V, the linear structure disintegrates into complete filling of the phase-space between a lower and an upper bound.  On a Poincar\'{e} plot, this is clear evidence of chaotic behavior.  The space filling is a graphic representation of the fact that even when subjected to a periodic potential, the response of the system at later times is unpredictable (within system bounds, as well as restrictions of harmonic driving).  There is some interesting fine-structure within part of the chaotic-memristor Poincar\'{e} plot, including one sub-voltage peak at about 0.8 V,  and the existence of what appears to be a tri-stable region near 1.25 V. This is typical of chaotic systems (see, e.g., Ref.~\cite{Mike2003}).

\section{Theoretical analysis}
\label{sec:3}

\begin{figure}[tbh] 
\centering
\includegraphics[width=7cm]{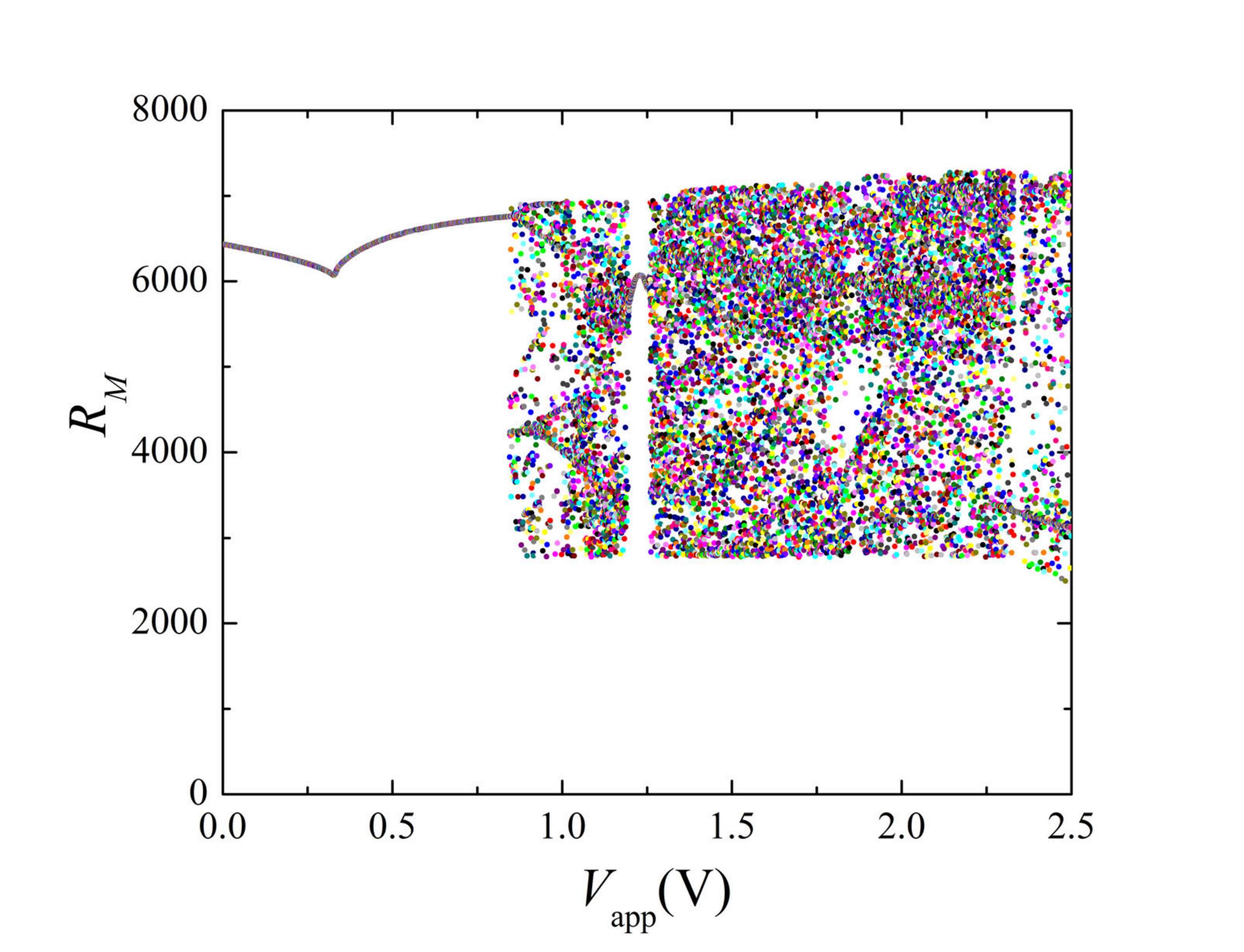}
\caption{Numerically calculated value of memristance $R_M(t_f)$ at
$t_f=19.875$s.  This is plotted as a function of the amplitude of the applied voltage $V_{app}$.  The fact that small changes in $V_{app}$ (i.e., neighboring x-axis points) result in large changes in $R_M(t_f)$ is an indicator of chaos. }
\label{FigRV}
\end{figure}

One simple way to understand the chaotic nature of the response seen in Fig.~\ref{FigPoincare} is to look at the instantaneous resistance value of the memristor after some fixed time.  Because the memristance value is so sensitive to the complete history of applied voltage, memristors with certain state-response functions - including, but likely not limited to, the double-well function used in this work - can enter a chaotic region where a small change in initial conditions produces a radically different value after a fixed time; the so called ``butterfly effect".  We can illustrate this quite nicely numerically.  To do so, we repeat the procedure of our experiment, driving the system with sinusoidal $V(t)$ and solving for a final value $R_M(t_f)$ at $t_f=19.875$ seconds.  In a non-chaotic system, small changes in the driving voltage amplitude $V_{app}$ would cause small changes in $R_M(t_f)$.  In our chaotic memristive system, however, small changes produce a pseudo-random value of $R_M(t_f)$, within certain bounds.  This is shown in Fig. \ref{FigRV}.  One advantage of illustrating this numerically is that it eliminates noise or other experimental errors from affecting the long-time value of $R_M(t_f)$, which can give the false impression of chaotic behavior even when none exists.

\begin{figure}[tbh] 
\centering
\includegraphics[width=7cm]{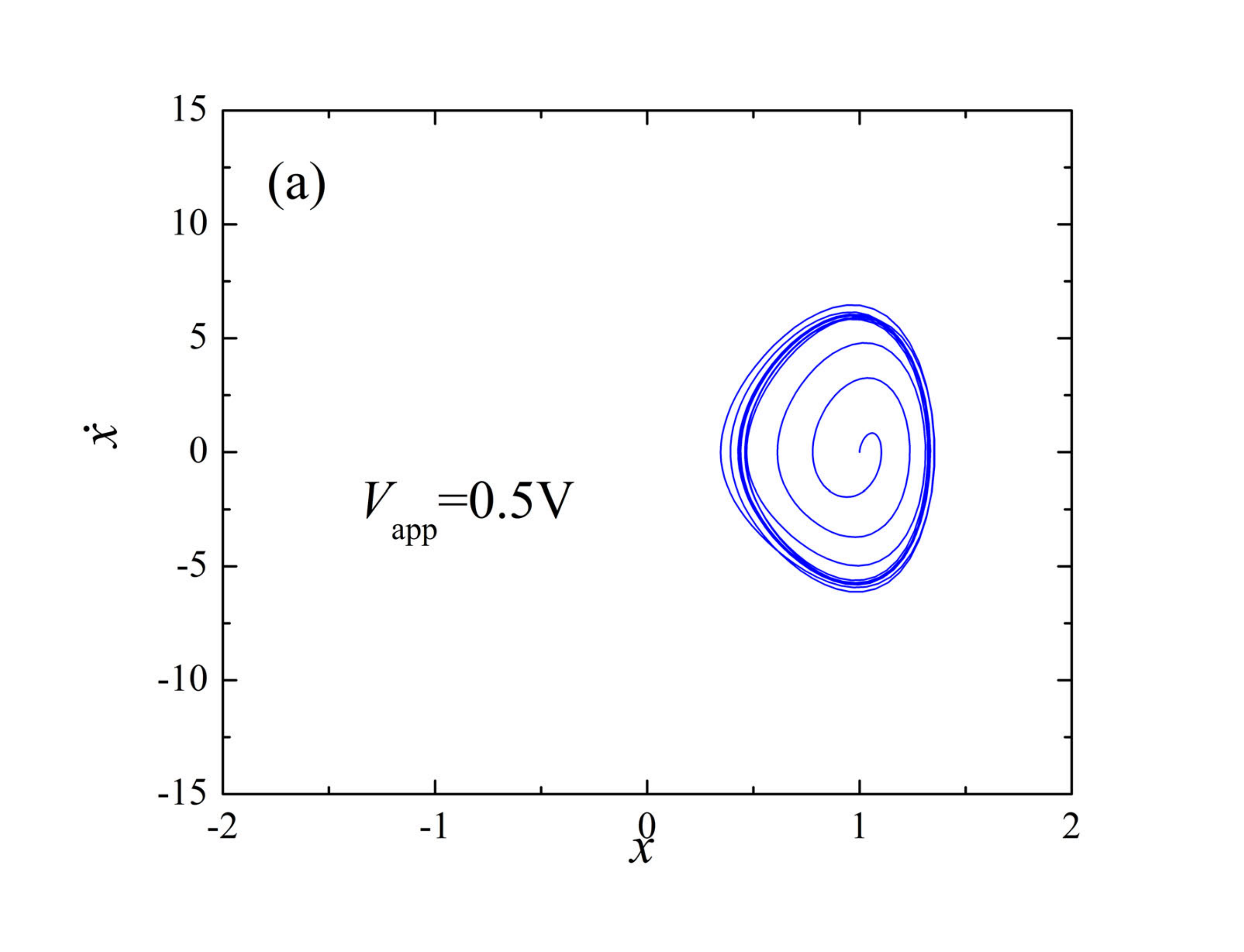}
\includegraphics[width=7cm]{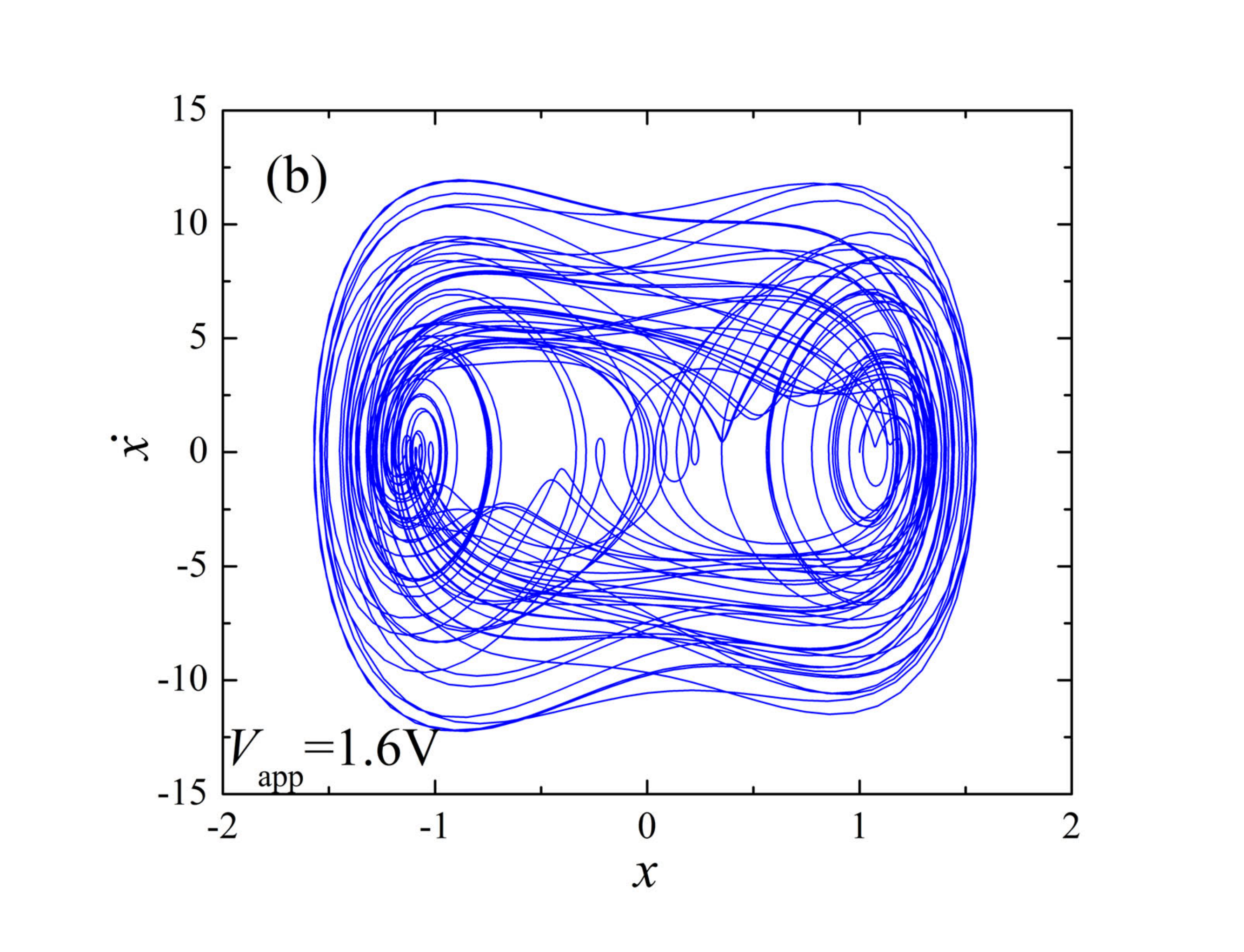}
\caption{System trajectories below (a) and above (b) the chaos onset. These plots represent 50s of system evolution driven by an
applied ac voltage of 2 Hz frequency. The voltage amplitude is shown on the plots for each case. }
\label{FigPhase}
\end{figure}

One final theoretical tool which is often employed to help understand chaotic systems is the phase-space attractor plot and accompanying calculation of the Lyapunov exponents.  For the double-well function of Eq. (\ref{programmed_eq}) which we use, the phase-space can be plotted as $\dot{x}(x)$.  This is shown in Fig. \ref{FigPhase} for two voltages, one below threshold (0.5V) and one above (1.6V).  Below threshold (Fig \ref{FigPhase}a) we observe the phase-space trajectory quickly moves from initial conditions into a steady orbit.  This is non-chaotic behavior.  Above threshold (Fig \ref{FigPhase}b), there is no single steady orbit, but instead the system circles around two attractors and explores almost an entire available phase space (determined by both both $V_{Rm}$ and system properties).

The Lyapunov exponent is a measure of how quickly two orbits in the phase space, originating from infinitesimally close initial conditions, will diverge in time, as expressed in the equation:
\begin{equation}\label{Lyapunov}
\vert \delta R_M(t) \vert \approx e^{\lambda t} \vert \delta R_M(t_0) \vert
\end{equation}
We calculate the maximum Lyapunov exponent for our system using the common numerical method described in \cite{Wolf85,Gallas95}.  For the two voltages in Fig. \ref{FigPhase}a,b (above and below threshold), we obtain values for the maximum Lyapunov exponent of $\lambda=-2.1 s^{-1} $ and $\lambda=0.15 s^{-1} $, respectively.  As the latter (above-threshold) value is greater than zero, it  proves the chaotic nature of the double-well system above threshold volage.  Although it would be gratifying to see a similar result obtained from our experimentally collected data, noise and limited data make this difficult.  This is not surprising, as unambiguous Lyapunov exponent calculation from data is often a difficult matter.

\section{Summary and conclusions}
\label{sec:4}

In this work, we have proposed a model of chaotic memristor whose internal state dynamics is described by an equation of a particle in multi-well potential. As an example of such a model, a nonauthonomous chaotic circuit based on a memristor governed by a double-well potential was investigated both experimentally and theoretically.  This simple circuit consists only of a memristor and a series current-sensing resistor.  Using a memristor emulator, we have experimentally demonstrated the chaotic operation of this circuit, and contrasted it to the observed non-chaotic response when a traditional control memristor is used.  By subjecting our chaotic memristive circuit to periodic applied voltages, we have constructed a Poincar\'{e} map for our system which clearly identifies a region of chaotic behavior above a threshold voltage.  The simplicity of this chaotic circuit should entice opportunities in many applications. Also, awareness of the existence of a chaotic region which may potentially exist in many memristive systems is critical for avoiding (or utilizing) this region as necessary.

Moreover, the similarity between our model of chaotic memristor and a wide range of physical systems, including TiO$_2$ memristors \cite{yang08a}, suggests that our results can be realizable in certain material systems. Indeed, the oxygen vacancies' migration in TiO$_2$ memristors can be considered as a hopping of a classical particle between potential minima \cite{strukov09a}. Such a process, for a single vacancy, is exactly described by Eq. (\ref{eqn_motion}), where $x$ is the vacancy's position and $U(x)$ is the potential energy profile along the vacancy's path. As the vacancies' positions define the transport properties of memristive devices \cite{yang08a}, chaotic features of vacancies motion should be visible in $I-V$ response. In fact, the experimentally observed noise in periodically driven TiO$_2$ memristors (see, for example, Fig. 1c of Ref. \cite{yang08a}) may be related to chaotic dynamics of oxygen vacancies.

\section*{Acknowledgments}
This work was supported by the Air Force Office of Scientific Research and by the National Science Foundation (DMR-0802830).  One of the authors (TD) would like to acknowledge support by the IC postdoctoral fellowship program.


\bibliography{memchaos2}

\end{document}